\documentclass[12pt,english,longbibliography,nofootinbib,superscriptaddress,12pt,sort&compress,showkeys]{extarticle}
\usepackage[T1]{fontenc}
\usepackage[latin9]{inputenc}
\usepackage{xcolor}
\usepackage{amsmath}
\usepackage{amssymb}
\usepackage{esint}
\usepackage[numbers]{natbib}
\PassOptionsToPackage{normalem}{ulem}
\usepackage{ulem}

\makeatletter

\newcommand{\lyxmathsym}[1]{\ifmmode\begingroup\def\b@ld{bold}
  \text{\ifx\math@version\bold\bfseries\fi#1}\endgroup\else#1\fi}

\providecommand{\tabularnewline}{\\}
\providecolor{lyxadded}{rgb}{0,0,1}
\providecolor{lyxdeleted}{rgb}{1,0,0}

\usepackage{indentfirst}
\usepackage{amsfonts}
\usepackage[T1]{fontenc}
\usepackage{ae,aecompl}
\usepackage{sidecap}
\usepackage[section]{placeins}
\usepackage{epsf}
\usepackage{graphicx}
\makeatother

\usepackage{babel}
\date{\today}

\usepackage{chngcntr}

\makeatother

\usepackage{babel}
\begin{document}
\title{Interatomic potentials for platinum\textbf{ }}
\author{R. K. Koju$^{1}$, Y. Li$^{1}$, and Y. Mishin$^{1}$\footnote{Corresponding author. Email: ymishin@gmu.edu}\\
$^{1}${\normalsize Department of Physics and Astronomy, MSN 3F3},\\
 {\normalsize George Mason University, Fairfax, VA 22030, USA}\\
}

\maketitle
\begin{abstract}
 We present two new interatomic potentials for platinum (Pt) in angular-dependent potential (ADP) and modified Tersoff (MT) formats. Both potentials have been trained on a reference database of first-principles calculations without using experimental data. The properties of Pt predicted by the ADP and MT potentials agree better with DFT calculations and experimental data than the potentials available in the literature. Future applications of the MT model to mixed-bonding metal-covalent systems are discussed.   
\end{abstract}

\date{}

\bigskip{}


\section{Introduction\label{sec:Introduction}}

Platinum (Pt) is a noble metal that favorably combines mechanical strength and relatively high melting temperature with exceptionally good oxidation resistance up to high temperatures. Pt has many technological applications in areas such as catalysis, metallic interconnects in microelectronic devices, and medicine. 

Classical atomistic simulations can provide valuable insights into the mechanical behavior and microstructural development in Pt, including its responses to high temperatures and  pressures \cite{Padilla-Espinosa:2022aa,Ding:2022aa,Azadehranjbar:2023aa}. The reliability of the results obtained from the atomistic simulations depends on the accuracy of the interatomic potential. Most of the potentials for Pt have been developed in the embedded-atom method (EAM) format \cite{Daw83}. To assess their accuracy, we have tested several recently published EAM potentials. Based on the testing results, we selected two most recent EAM potentials that we consider to be more reliable than others: the EAM potential developed by Zhou et al.~\cite{Zhou2004a} and the EAM potential developed by O'Brien et al.~\cite{OBrien:2018aa}. We will refer to these potentials as EAM1 and EAM2, respectively. For comparison, Table \ref{tab:other_EAM} in the Appendix to this paper summarizes properties computed with two earlier EAM potentials \cite{Adams89a,Foiles86}. For completeness, Table \ref{tab:other_EAM} also includes an effective medium theory (EMT) potential \cite{Jacobsen1996} and a modified EAM (MEAM) potential \cite{Lee03a}. All potentials tested here exhibit several limitations, including significantly underestimated stacking-fault energies and/or  melting temperatures that are several hundred degrees below or above the experimental values. Such deficiencies can reduce the quantitative accuracy of simulations and, in some cases, result in nonphysical behavior. 

To address the limited accuracy of the existing Pt potentials, we have developed two new interatomic potentials for Pt. One potential has the angular dependent potential (ADP) format \citep{Mishin05a,Mishin.Ta,Hashibon08,Apostol2010,Apostol2011}, which is an extension of EAM to include angular dependent interactions. The other potential has the modified Tersoff (MT) format and is a relatively recent development. The original Tersoff potentials \citep{Tersoff88,Tersoff:1988dn,Tersoff:1989wj} were designed for strongly covalent elements, such as Si and C. Their cutoff radius is usually chosen between the first and second coordination shells of the diamond cubic structure. Here, we apply the MT model to a metal (Pt) by expanding the cutoff radius to include several coordination shells. In this form, MT becomes a many-body atomic interaction model similar to the modified EAM (MEAM) \cite{Baskes87} and ADP. Both the ADP and MT potentials were trained on a large database of first-principles calculations and tested for their ability to reproduce a broad range of Pt properties.

In section \ref{subsec:DFT-calculations}, we present the methodology of the first-principles calculations, which were used to generate the training database  and compute Pt properties for comparison with potential predictions. Section \ref{sec:MD} presents our methodology for computing the properties of Pt using the new potentials. For consistency, we have applied the same methodology to recompute all properties with the EAM1 and EAM2 potentials. In section \ref{sec:Potential-format}, we review the potential formats and describe the fitting procedures. The central part of the paper is section \ref{sec:properties}, in which we report the properties of Pt computed with the new potentials, as well as the EAM1 and EAM2 potentials, and compare them with the DFT calculations and experimental data. The strengths and weaknesses of the potentials are discussed in detail. The comparison demonstrates an improved accuracy of the new potentials relative to the existing potentials.  Finally, in section \ref{sec:conclusions}, we summarize the work and put it in perspective with the potential development field.

\section{First-principles calculations\label{subsec:DFT-calculations}}

\subsection{Methodology}

The first-principles calculations were based on density functional theory (DFT) and used the Vienna \textit{Ab initio} Simulation Package (VASP) \citep{Kresse1996,Kresse1996b}.
We used the projector-augmented wave (PAW) method \citep{Blochl94}
and the generalized PBE (Perdew-Burke-Ernzerhof) gradient approximation
for exchange--correlation \citep{PerdewBE96}. The PBE data set for
Pt was \texttt{PAW\_PBE} 05Jan2001 with the valence electron configuration
s$^{1}$d$^{9}$. Methfessel--Paxton smearing with a width of 0.05 eV was applied
during the Brillouin-zone integration. The tolerance used for the
ionic and electronic relaxations was $10^{-7}\,$ eV and $10^{-8}\,$
eV, respectively. The ionic relaxation was considered to have converged when
the maximum residual force was below $10^{-3}\,\mathrm{eV}\,\text{\AA}^{-1}$. 

The DFT calculations served two purposes: to generate a reference
database for training the classical potentials and to calculate the physical properties of Pt for comparison with predictions of the potentials.

\subsection{The reference database}

For each crystal structure included in the database, energy convergence
tests were performed to find the optimal energy cutoff (520 eV) and k-point
mesh. The energy-volume relations for the crystal structures, which
we refer to as the equations of states (EOS), were calculated without
ionic relaxation using the tetrahedron method with Bl\"ochl corrections
(ISMEAR=$-5$). For calculating the vacancy migration path and surface energies, the partial occupancies for each orbital were changed to 1 (ISMEAR=1).
Instead of running ab-initio molecular dynamics simulations, we used
surface and liquid structures generated with a classical potential
\citep{Zhou2004a} as input to VASP and evaluated their energies
and forces without relaxation. 

The reference database included a set of EOSs for FCC, body-centered
cubic (BCC), hexagonal close-packed (HCP), simple cubic (SC), A15, 
and diamond cubic (DC) structures under isotropic deformations. The
FCC structure was additionally subjected to uniaxial tension and compression
along the $\left\langle 100\right\rangle $, $\left\langle 110\right\rangle $ 
and $\left\langle 111\right\rangle $ directions, as well as to shear
deformations. The shear deformations were designed to represent the elastic moduli $(C_{11}-C_{12})$ and $C_{44}$. The database also included the minimum energy path (MEP)
for a vacancy jump computed by the nudged elastic band (NEB) method \cite{Jonsson98,HenkelmanJ00}
within VASP using nine images of a primitive supercell $3\times3\times3$.
In addition, an MEP containing more images was computed with a classical
potential \citep{Zhou2004a} and the supercell energies and atomic
force were recomputed with VASP. 

Before using the database to train potentials, all DFT energies were shifted by subtracting the energy of an isolated atom computed with VASP ($-0.17$ eV/atom). This made the database consistent with the calibration of the interatomic potentials, for which the energy of an isolated atom is set to zero. This shift did not affect the defect energies. Table \ref{table:DFT-database} provides
a more detailed description of the DFT database. 

\subsection{Calculation of properties}

In the DFT calculations of Pt properties, Brillouin-zone integrations were performed using Monkhorst--Pack meshes with a density of $10{,}000$ k-points per reciprocal atom (kppa $= 10{,}000$). All structural relaxations were performed using the conjugate-gradient algorithm, allowing full relaxation of atomic positions as well as cell volume and shape. The energy and force convergence criteria were specified in Section 2.1.

The equilibrium lattice constants and the cohesive energies of the crystalline structures were obtained by fully relaxing a $3 \times 3 \times 3$ supercell. The elastic constants were calculated by applying small strains $\epsilon$ (ranging from $-0.003$ to $0.003$ in increments of $0.001$) to a well-relaxed $3 \times 3 \times 3$ FCC supercell along different crystallographic orientations and computing the corresponding stress $\sigma$ as a function of $\epsilon$. The resulting stress--strain relations were linearly fitted and the elastic constants $C_{ij}$ were obtained from the slope according to the relation $C_{ij} = d\sigma / d\varepsilon$.

Point-defect energies were calculated by removing one atom to create a vacancy or adding one atom to form an interstitial in a fully relaxed $3 \times 3 \times 3$ FCC supercell. The interstitial configurations considered here included tetrahedral, octahedral, and dumbbell, with dumbbell orientations along the $\langle100\rangle$, $\langle110\rangle$, and $\langle111\rangle$ directions. The point-defect formation energy $E_f$ was determined as the energy difference between a relaxed supercell containing the defect and the corresponding perfect-lattice supercell, expressed as
\begin{equation}
E_f = E_d - \frac{N \pm 1}{N} E_p ,
\end{equation}
where $E_d$ is the total energy of the relaxed supercell containing the point defect and $N \pm 1$ atoms, and $E_p$ is the total energy of a perfect-lattice supercell containing $N$ atoms. The migration energy barrier of a vacancy was calculated using the NEB method \cite{Jonsson98,HenkelmanJ00} implemented in VASP.

Surface energies were computed using slab models with periodic boundary conditions in all three directions. A vacuum layer of more than 1~nm was introduced along the $z$ direction to generate two free surfaces. A similar supercell without vacuum was used as a reference. The separation distance between the two surfaces, corresponding to the thickness of the slab along the direction $z$, was at least 1.9~nm to eliminate spurious interactions between periodic images. The system size ranged from 24 to 72 atoms, depending on the specific surface orientation. The surface energy was calculated from the equation
\begin{equation}
\gamma_s = \frac{E_s - E_p}{2\, l_x l_y},
\end{equation}
where $E_s$ is the total energy of the relaxed supercell containing two free surfaces, $E_p$ is the total energy of the reference perfect-lattice supercell, and $l_x$ and $l_y$ are the lateral dimensions of the surface.

The twin boundary energy was computed by relaxing a supercell that contained a coherent twin boundary, with dimensions of $5.5~\text{\AA} \times 19.2~\text{\AA} \times 54.3~\text{\AA}$. 
The twin boundary was constructed by rotating the upper half of the crystal by $180^\circ$ about an axis normal to the $(111)$ plane. 
The twin boundary energy was calculated as the energy difference between the relaxed supercell containing the twin boundary and the corresponding perfect-lattice supercell of the same size, normalized by the twin boundary area. Because the periodic supercell contained two equivalent twin boundaries, the result was divided by two.

The $\gamma$-surface on a (111) plane was computed using the tilted-cell method~\cite{kibey2006generalized}. A fully relaxed, defect-free supercell containing twelve \((111)\) atomic layers was sheared by introducing a $YZ$ tilt of magnitude $u$, corresponding to a displacement along $[11\bar{2}]$, while keeping the Cartesian coordinates of all atoms fixed (i.e., no remapping of atomic positions with cell deformation). Periodic boundary conditions were applied in all directions, leading to the formation of a stacking fault at the periodic boundary. With the shape of the cell kept fixed, the atomic positions were relaxed only in the $[111]$ direction. The $\gamma$-surface was obtained from the energy difference between the faulted and perfect-lattice configurations normalized by the interfacial area of the \((111)\) plane.

For the Pt dimer Pt$_2$, a system consisting of two atoms separated by a distance of $4~\text{\AA}$ was constructed. A supercell of size $20~\text{\AA} \times 20~\text{\AA} \times 20~\text{\AA}$ was used to minimize boundary effects. The system was then fully relaxed, and the dimer energy was obtained by calculating the energy difference per atom between the relaxed dimer configuration and two isolated atoms.

The phonon calculations utilized the \textsf{phonopy} software package \citep{phonopy}. An atom in a primitive supercell $4\times4\times4$  was displaced by 0.01 $\textrm{\AA}$, and the energy and forces calculated
by VASP were used as input in \textsf{phonopy} to evaluate 
the density of states (DOS) and the phonon dispersion curves.

\section{Calculations with interatomic potentials\label{sec:MD}}

We used the molecular dynamics code \textsc{LAMMPS}~\cite{Plimpton95} to calculate the properties of  Pt with the interatomic potentials developed in this work and available in the literature~\cite{Zhou2004a, OBrien:2018aa}. 

The equilibrium lattice constants and the cohesive energies of the crystal structures were obtained by anisotropic relaxation of a $5 \times 5 \times 5$ supercell. The EOSs of the  crystal structures were computed by volumetric deformation of the relaxed structures to different atomic volumes and calculation of the energy per atom. The elastic constants of Pt were calculated by applying small strains $\epsilon$ of $\pm 1 \times 10^{-6}$ to a relaxed cubic FCC supercell, computing the resulting stress tensors, and evaluating the elastic response by numerical differentiation: $C_{ij} = d\sigma / d\epsilon$.

Point-defect energies were calculated by removing one atom to create a vacancy or adding one atom to create an interstitial defect in a well-relaxed $5 \times 5 \times 5$ FCC supercell. The approach was similar to that used in the DFT calculations. After relaxation, the point defect formation energy $E_f$ was calculated as
\begin{equation}
E_f = E_d - (N \pm 1) E_0 ,
\end{equation}
where $E_d$ is the total energy of the relaxed supercell containing the point defect, $N$ is the number of atoms in the perfect-lattice supercell, and $E_0$ is the cohesive energy per atom of the perfect crystal. The vacancy migration energy barrier was calculated using the NEB method \cite{Jonsson98,HenkelmanJ00} implemented in \textsc{LAMMPS}.

The surface energy was evaluated by relaxing a simulation slab containing two free surfaces, with periodic boundary conditions applied along the directions  $x$ and $y$  parallel to the surfaces.
The separation distance between the two surfaces was at least 2~nm to eliminate spurious interactions. The system was relaxed using energy minimization.
The surface energy was calculated as
\begin{equation}
\gamma_s = \frac{E_{s} - N E_0}{2 l_x l_y} ,
\end{equation}
where $E_{s}$ is the total energy of the relaxed supercell containing two surfaces, $N$ is the number of atoms in the supercell, and $l_x$ and $l_y$ are the lateral dimensions of the surface.

The twin boundary energy was calculated by relaxing a simulation block containing a coherent twin boundary, with dimensions of $34~\text{\AA} \times 59~\text{\AA} \times 84~\text{\AA}$. 
The twin boundary was constructed by rotating the upper half of the crystal by  $180^\circ$ about the $[111]$ axis. 
The twin boundary energy was obtained as the total energy of the relaxed supercell containing the twin boundaries minus the number of atoms multiplied by $E_0$ and normalized by the twin boundary area. 
Because the periodic supercell contained two equivalent twin boundaries, the result was divided by two.

For $\gamma$-surface calculations, an FCC crystal was constructed with crystallographic orientations
$x \parallel [1\bar{1}0]$, $y \parallel [11\bar{2}]$, and $z \parallel [111]$. 
Periodic boundary conditions were applied along the $x$ and $y$ directions, while free boundary conditions were imposed along the $z$ direction. 
The dimensions of the simulation cell were approximately $3.4~\mathrm{nm} \times 3.9~\mathrm{nm} \times 9.6~\mathrm{nm}$. The $\gamma$-surface was generated by rigidly displacing the upper half of the crystal relative to the lower half along the $[11\bar{2}]$ direction in the $(111)$ plane. 
After each displacement, the system was relaxed by energy minimization, with atomic relaxation restricted to the  $z$  direction normal to the fault plane. 
The energy change relative to the  perfect-lattice supercell was calculated after each displacement and divided by the cross-sectional area of the fault (i.e., the $xy$-plane area of the simulation cell) to obtain the fault energy $\gamma$.

For a dimer Pt$_2$, the calculation procedure was similar to that used in the DFT calculations. A system consisting of two atoms separated by a distance of $4~\text{\AA}$ was constructed using non-periodic boundary conditions. The system was then fully relaxed, and the dimer energy was obtained by calculating half of the total energy.

The phonon calculations were performed using the \textsf{phonopy} software package \cite{phonopy}.  Atomic displacements were applied in a $8 \times 8 \times 8$ supercell at $0~\mathrm{K}$, and the resulting atomic forces were calculated using \textsc{LAMMPS}. 
The force snapshots were then used as input to compute the force constants and evaluate the phonon DOS and phonon dispersion relations.

The relationship between atomic volume and pressure was obtained by isotropically deforming a relaxed $5 \times 5 \times 5$ supercell. For each volume, the system was equilibrated under the NVT ensemble at room temperature for 40~ps. The pressure corresponding to each volume was calculated by averaging the system pressure over the last 20~ps of the simulation.

The linear thermal expansion coefficient at each temperature was determined by averaging the dimensions of a $6 \times 6 \times 6$ supercell equilibrated in the $NPT$ ensemble at the target temperature for $80~\mathrm{ps}$. The average dimensions of the supercell were then compared with those at $300~\mathrm{K}$ to compute the relative thermal expansion.

The melting temperature was calculated using the phase coexistence method~\cite{Morris94, Morris02, Howells:2018aa}. 
Specifically, a periodic simulation cell was constructed with crystallographic orientations 
$x \parallel [100]$, $y \parallel [010]$, and $z \parallel [001]$. 
The simulation cell had dimensions of $2 \times 2 \times 40~\mathrm{nm}^3$ and contained approximately $10{,}000$ atoms. The system was first equilibrated in the $NPT$ ensemble at an initial temperature close to the expected melting point. 
Subsequently, one half of the simulation cell was heated to a sufficiently high temperature to cause melting. The  system obtained contained coexisting solid and liquid regions separated by a solid--liquid interface parallel to the $xy$ plane. The system was  annealed in the $NPH$ ensemble under an external pressure of $1~\mathrm{atm}$ applied along the normal direction to the interface ($z$ direction). As the simulation progressed, partial melting or solidification occurred, leading to a transient temperature variation. 
This temperature drift gradually diminished as the system approached the solid--liquid phase equilibrium. 
Once equilibrium was established, the temperature stabilized at $T_m$, which was identified as the melting temperature at ambient pressure. The equilibrium temperature was averaged over several sequential 20 ps intervals, and these values were used to compute the final $T_m$ and its standard deviation.

\section{Potential formats and fitting procedures\label{sec:Potential-format}}

\subsection{The angular-dependent potential}

In the ADP formalism \citep{Mishin05a,Mishin.Ta,Hashibon08,Apostol2010,Apostol2011},
the total energy of a single-component system is expressed by 
\begin{equation}
E_{tot}=\frac{1}{2}\sum_{i,j(i\ne j)}V(r_{ij})+\sum_{i}F(\bar{\rho}_{i})+\frac{1}{2}\sum_{i,\alpha}(\mu_{i}^{\alpha})^{2}+\frac{1}{2}\sum_{i,\alpha,\beta}(\lambda_{i}^{\alpha\beta})^{2}-\frac{1}{6}\sum_{i}\nu_{i}^{2},\label{eqn:eos}
\end{equation}
where the indices $i$ and $j$ enumerate the atoms, and the superscripts $\alpha,\beta=1,2,3$
represent the Cartesian components of the vectors and tensors. In Eq.\,(\ref{eqn:eos}),
$V(r_{ij})$ is the pair interaction energy between an atom $i$ located
at position $\textsl{\textbf{r}}_{i}$ and an atom $j$ located at
position $\textsl{\textbf{r}}_{j}=\textsl{\textbf{r}}_{i}+\textsl{\textbf{r}}_{ij}$.
The second term in Eq.\,(\ref{eqn:eos}) is the energy of embedding
an atom $i$ in the host electron density $\bar{\rho}_{i}$ induced
by all other atoms. The host electron density is calculated by 
\begin{equation}
\bar{\rho}_{i}=\sum_{j\ne i}\rho(r_{ij}),\label{eqn:electron}
\end{equation}
where $\rho(r_{ij})$ is the electron density function of a
neighboring atom $j$. The first two terms in Eq.\,(\ref{eqn:eos})
constitute the standard EAM format \citep{Daw83}. The remaining terms
in Eq.\,(\ref{eqn:eos}) reflect the non-central character of bondings
and include the dipole vectors 
\begin{equation}
\mu_{i}^{\alpha}=\sum_{j\ne i}u(r_{ij})r_{ij}^{\alpha},\label{eqn:dipole}
\end{equation}
and quadrupole tensors 
\begin{equation}
\lambda_{i}^{\alpha\beta}=\sum_{j\ne i}w(r_{ij})r_{ij}^{\alpha}r_{ij}^{\beta},\label{eqn:quadruple}
\end{equation}
where the trace of $\lambda_{i}^{\alpha\beta}$ is 
\begin{equation}
\nu_{i}=\sum_{\alpha}\lambda_{i}^{\alpha\alpha}.\label{eqn:trace}
\end{equation}
In these equations, $u(r_{ij})$ and $w(r_{ij})$ are additional potential functions
of the radial distance $r_{ij}$ between atoms. 

For Pt, the pair-interaction function $V(r)$ was parametrized in the
form
\begin{equation}
V(r)=\left[M(r;E_{1},r_{01},b_{1})+M(r;E_{2},r_{02},b_{2})+s_{1}\right]f_{c}\left(\dfrac{r-r_{c}}{h}\right)+m\rho(r),\label{eq:pair}
\end{equation}
where 
\[
M(r;E,r_{0},b)=E\left[\left(e^{-b(r-r_{0})}-1\right)^{2}-1\right]
\]
is the Morse function, and
\begin{equation}
f_{c}(x)=\begin{cases}
{\displaystyle \dfrac{x^{4}}{1+x^{4}}} & \textrm{if \ensuremath{x<0},}\\
0 & \textrm{if \ensuremath{x\geq0}}
\end{cases}\label{eq:cutoff-func-1}
\end{equation}
is a cutoff function. Note that in Eq.(\ref{eq:pair}), we mix the
functions $V(r)$ and $\rho(r)$ with the mixing factor $m$ to enhance
the flexibility of the shape. The functions $\rho(r)$, $u(r)$, and $w(r)$
were chosen in the same functional form, but with different parameters:
\[
\rho(r)=\left[G(r;E_{3},r_{03},b_{3},0)+G(r;E_{4},r_{04},b_{4},2)+s_{2}\right]f_{c}\left(\dfrac{r-r_{c}}{h}\right),
\]
\[
u(r)=\left[G(r;E_{5},r_{05},b_{5},0)+G(r;E_{6},r_{06},b_{6},2)+s_{3}\right]f_{c}\left(\dfrac{r-r_{c}}{h}\right),
\]
\[
w(r)=\left[G(r;E_{7},r_{07},b_{7},0)+G(r;E_{8},r_{08},b_{8},2)+s_{4}\right]f_{c}\left(\dfrac{r-r_{c}}{h}\right),
\]
where
\[
G(r;E,r_{0},b,n)=Er^{n}e^{-b(r-r_{0})^{2}}.
\]
The embedding energy $F$ as a function of the host electron density
$\overline{\rho}$ had the functional form
\[
F(\overline{\rho})=\rho_{0}\dfrac{\overline{\rho}+a\overline{\rho}^{2}+b\overline{\rho}^{3}+c\overline{\rho}^{4}}{1+d\overline{\rho}^{3}}.
\]
This function satisfied the boundary conditions $F(0)=0$ and $F\rightarrow\rho_{0}(c/d)\overline{\rho}$
at $\overline{\rho}\rightarrow\infty$.

The potential has a total of 36 fitting parameters. They were obtained
by minimizing the weighted mean-square deviation of supercell 
energies from the reference DFT data using a simulated annealing algorithm
based on the Nelder-Mead simplex method \citep{Nelder:1965aa}. The
optimized parameters are listed in Table \ref{tab:Optimized-parameters-ADP}. Supercell weights were used to control the relative accuracy of different physical properties predicted by the potential. In practice, multiple optimization runs were performed, followed by testing the properties and placing greater weight on the supercells that represented the least accurate properties. The iterations in this feedback loop continued until a satisfactory combination of properties was reached. 

\subsection{The modified Tersoff potential\label{sec:MT}}

In the single-component MT model, the total energy is represented
in the form 
\[
E=\dfrac{1}{2}\sum_{i\neq j}V_{ij}(r_{ij}),
\]
where $r_{ij}$ is the distance between the atoms $i$ and $j$. The pair
interaction energy $V_{ij}$ has the form 
\begin{equation}
V_{ij}=\left[A\exp(-\lambda_{1}r_{ij})-b_{ij}B\exp(-\lambda_{2}r_{ij})+c_{0}\right]f_{c}(r_{ij}),\label{eq:MT_1}
\end{equation}
where the bond order factor $b_{ij}$ is given by 
\begin{equation}
b_{ij}=\left(1+\xi_{ij}^{\eta}\right)^{-\delta}.\label{eq:MT_2}
\end{equation}
Here, $\xi_{ij}$ is an angular-dependent three-body sum,
\begin{equation}
\xi_{ij}=\sum_{k\ne i,j}f_{c}(r_{ik})g(\theta_{ijk})\exp\left[\alpha(r_{ij}-r_{ik})^{\beta}\right],\label{eq:MT_3}
\end{equation}
where $\theta_{ijk}$ is the angle between the bonds $ij$ and $ik$.
Physically, $(1+\xi_{ij})$ represents an effective coordination number
of atom $i$. The cutoff function $f_{c}(r)$ has the form 
\[
f_{c}(r)=\begin{cases}
1, & r\le R_{1}\\
\dfrac{1}{2}+\dfrac{9}{16}\cos\left(\pi\dfrac{r-R_{1}}{R_{2}-R_{1}}\right)-\dfrac{1}{16}\cos\left(3\pi\dfrac{r-R_{1}}{R_{2}-R_{1}}\right), & R_{1}<r<R_{2}\\
0, & r\ge R_{2},
\end{cases}
\]
where $R_{1}$ and $R_{2}$ are the inner and outer cutoff radii.
The angular function $g(\theta)$ has the form 
\[
g(\theta)=c_{1}+\dfrac{c_{2}(h-\cos\theta)^{2}}{c_{3}+(h-\cos\theta)^{2}}\left\{ 1+c_{4}\exp\left[-c_{5}(h-\cos\theta)^{2}\right]\right\} .
\]

The original version of this potential was proposed by Tersoff \citep{Tersoff88,Tersoff:1988dn,Tersoff:1989wj}
and later modified by Kumagai et al.~\citep{Kumagai:2007ly} by generalizing
the angular function $g(\theta)$. More recently, it was proposed
\citep{Purja-Pun:2017aa} to add the coefficient $c_{0}$ to Eq.(\ref{eq:MT_1})
to better control the attractive part of the potential. The potential
has 16 free parameters: $A$, $B$, $\alpha$, $h$, $\eta$, $\lambda_{1}$,
$\lambda_{2}$, $R_{1}$, $R_{2}$, $\delta$, $c_{0}$, $c_{1}$,
$c_{2}$, $c_{3}$, $c_{4}$ and $c_{5}$. The power $\beta$ is a
fixed odd integer. In this work, we chose $\beta=1$. 

As noted in section \ref{sec:Introduction}, Tersoff potentials were
initially developed for strongly covalent elements. Consequently, the
outer cutoff $R_{2}$ was chosen between the first and second coordination
shells. Here, we apply the MT model to a metal by
expanding the outer cutoff $R_{2}$ to 6.57~$\textrm{\AA}$, which
includes several coordination shells of the FCC structure. Due to
this expansion, the MT model captures many-body interactions in a
manner similar to MEAM and ADP potentials, with $b_{ij}$ in Eq.(\ref{eq:MT_2})
playing the role of the embedding function and $\xi_{ij}$ being similar
to the electron density function. 

The MT potential was trained on the same DFT database and by the same
optimization method as the ADP potential. The optimized parameters
are reported in Table \ref{tab:Optimized-parameters-MT}.

\section{Platinum properties predicted by potentials\label{sec:properties}}

Table \ref{table:table1} summarizes the properties of Pt predicted by the new ADP and MT potentials. For comparison, the table includes predictions of the  EAM1 \cite{Zhou2004a} and EAM2 \cite{OBrien:2018aa} potentials, DFT calculations performed in this work, and experimental data when available. The EAM1 and EAM2 potential files were downloaded from the Interatomic Potentials Repository (https://www.ctcms.nist.gov/potentials/), and all properties predicted by these potentials were recomputed. 

\subsection{FCC lattice properties}
\begin{figure}[ht]
\vspace{0.0cm}
\centering
\includegraphics[height=5.9cm]{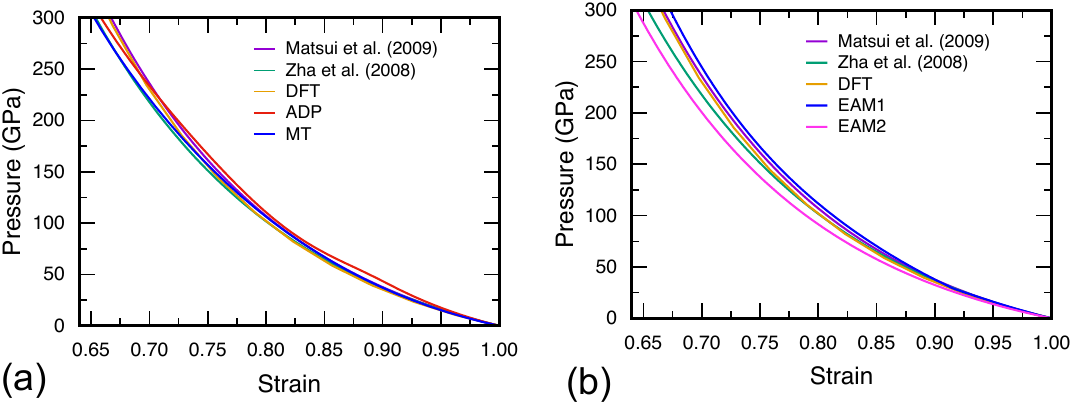}
\caption[]{Hydrostatic pressure as a function of volumetric strain for Pt obtained by experimental measurements by Matsui et al.~\cite{Matsui:2009aa} and Zha et al.~\cite{Zha:2008aa} in comparison with DFT calculations and predictions of interatomic potentials: (a) potentials developed in this work and (b) EAM1 \cite{Zhou2004a} and EAM2 \cite{OBrien:2018aa} potentials.}
\label{fig:pressure}
\end{figure}

\begin{figure}[ht]
\vspace{0.0cm}
\centering
\includegraphics[height=5.9cm]{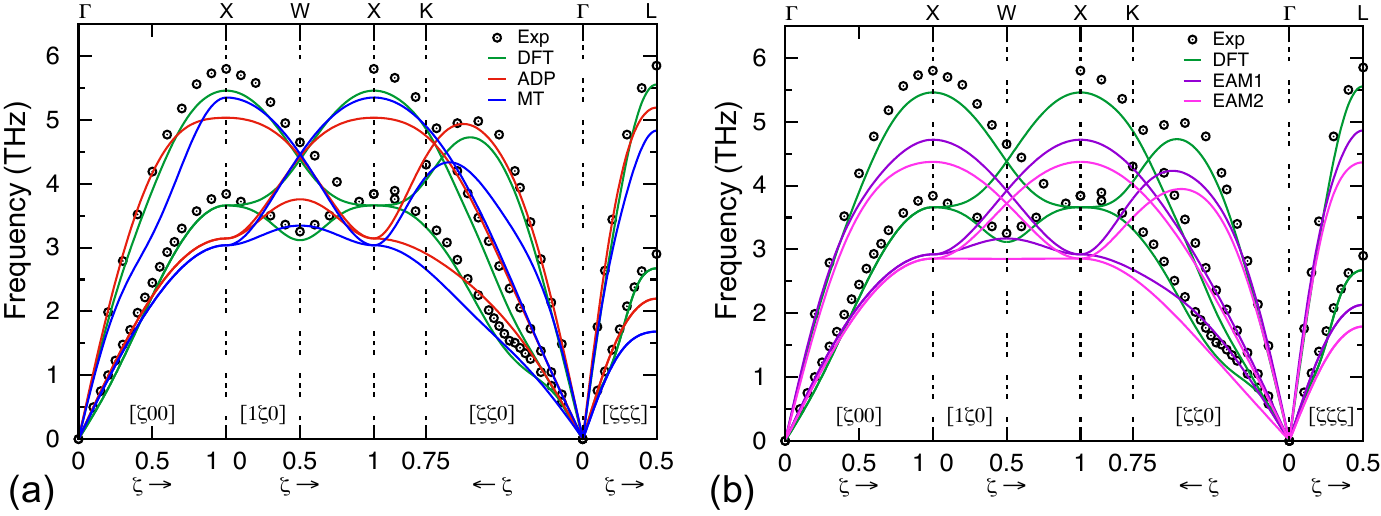}
\caption[]{Phonon dispersion relations for Pt obtained by experimental measurements \cite{dutton1972crystal} in comparison with DFT calculations and predictions of interatomic potentials: (a) potentials developed in this work and (b) EAM1 \cite{Zhou2004a} and EAM2 \cite{OBrien:2018aa} potentials.}
\label{fig:phonons}
\end{figure}

The equilibrium lattice parameter $a_0$, cohesive energy of the FCC structure $E_0$, and the elastic constants $C_{ij}$ are predicted by all four potentials reasonably well, although the ADP potential over-predicts $C_{44}$. Figure \ref{fig:pressure} compares the pressure-volume relations computed with the potentials together with experimental data and DFT calculations. The ADP and MT potentials compare with DFT and experimental data fairly well. The EAM1 and EAM2 potentials show more significant deviations. As another test of lattice properties, we computed the phonon dispersion relations and compared them with experimental data and DFT calculations. As shown in Fig.~\ref{fig:phonons}, the DFT calculations underestimate the experimental frequencies at the high-symmetry points X and L. The MT and ADP potentials underestimate the frequencies slightly more, and the EAM1 and EAM2 potentials even more. Overall, the new potentials perform significantly better in reproducing the phonon dispersion. Table \ref{table:table1} quantifies the differences by comparing the predicted maximum phonon frequencies $\nu_{\rm max}$ in the vibrational spectrum. 

\begin{figure}[ht]
\vspace{0.0cm}
\centering
\includegraphics[height=5.9cm]{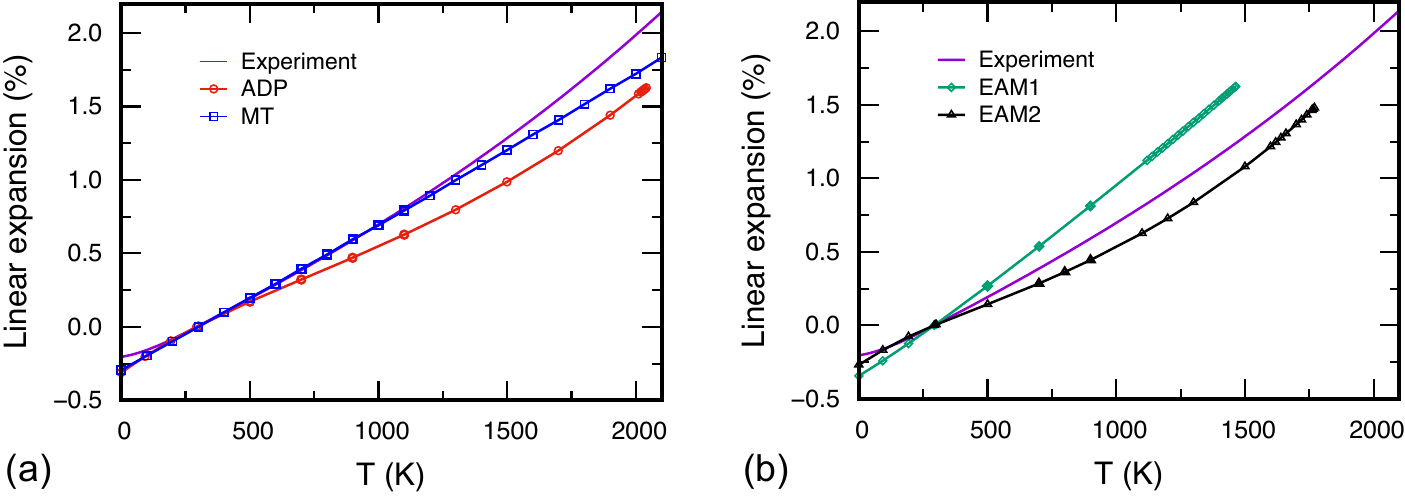}
\caption[]{Linear coefficient of thermal expansion of Pt relative to room temperature (293 K) obtained by experimental measurements \cite{Expansion} in comparison with DFT calculations and predictions of interatomic potentials: (a) potentials developed in this work and (b) EAM1 \cite{Zhou2004a} and EAM2 \cite{OBrien:2018aa} potentials. Each curve ends at the melting point predicted by the respective potential.}
\label{fig:thermal}
\end{figure}

To assess the anharmonicity of atomic vibrations, Fig.~\ref{fig:thermal} shows the linear thermal expansion factor (relative to room temperature) computed with the four potentials. The MT potential exhibits excellent agreement with the experimental temperature dependence of the expansion factor up to 1200 K, showing slight deviations at higher temperatures. The ADP potential is significantly less accurate, while the EAM1 and EAM2 potentials show even greater deviations from the experimental data. 

Another measure of the lattice stability is the melting temperature $T_m$. The potentials predict the  $T_m$ values of  $2041 \pm 5$ K (ADP), $2195 ± 6$ K (MT), $1474 \pm 1.3$ K (EAM1), and $1784 \pm 3$ K (EAM2), where the error bar represents one standard deviation. The ADP potential reproduces the experimental melting temperature within a few degrees (0.2 \%). The MT potential overestimates $T_m$ by 7.3 \%. In comparison, the EAM1 and EAM2 potentials underestimate the melting temperature by a few hundred K (Table \ref{table:table1}).

\subsection{Defect properties}
\begin{figure}[ht]
\vspace{0.0cm}
\centering
\includegraphics[height=5.5cm]{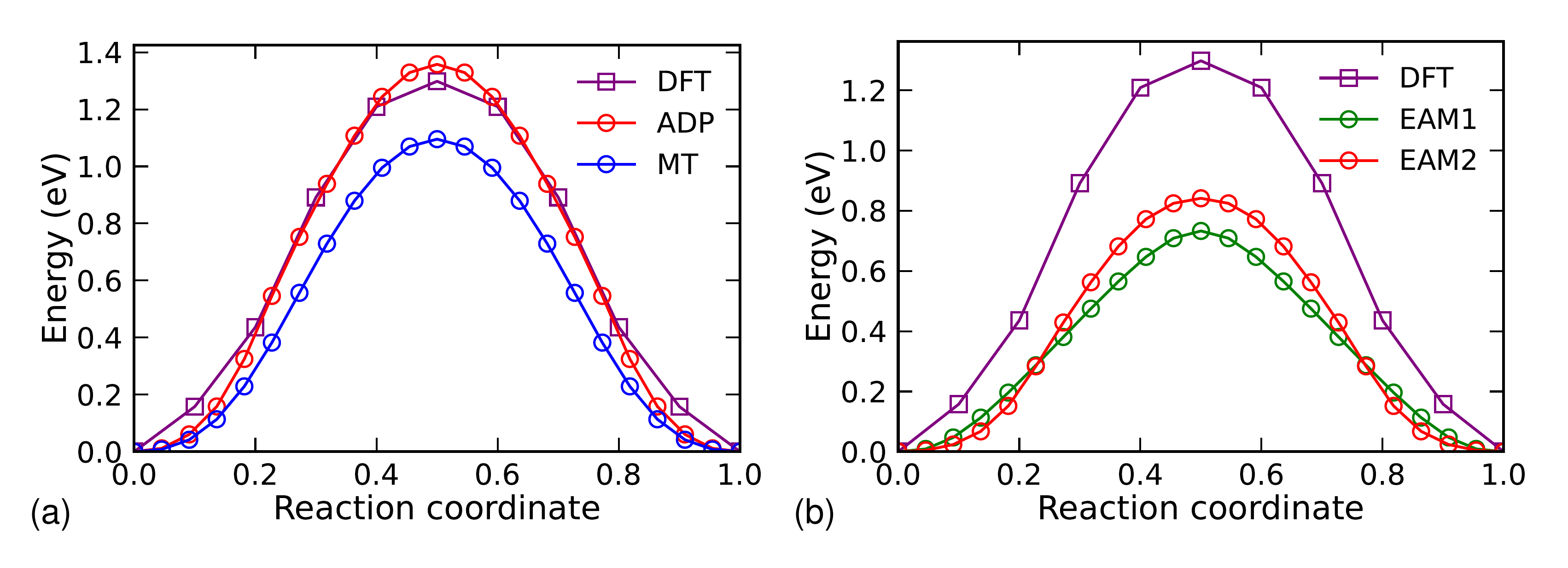}
\caption[]{Vacancy migration energy  in Pt obtained by DFT calculations  in comparison with (a) potentials developed in this work and (b) EAM1 \cite{Zhou2004a} and EAM2 \cite{OBrien:2018aa} potentials.}
\label{fig:neb}
\end{figure}

Table \ref{table:table1}  shows that the vacancy formation energy $E_f^v$ predicted by the potentials is within the range of the experimental data, while the DFT calculation yields an energy that is nearly half the experimental value. The gross underestimate of the vacancy formation energy by PBE DFT was previously reported by other authors \cite{Chapman:2020ti,OBrien:2018aa,Medasani:2015wu,Ma:2021aa}. This discrepancy is common to several other metals and is an intrinsic property of DFT functionals, as recently discussed in \cite{Medasani:2015wu}. Interestingly, the DFT value of the vacancy migration energy $E_m^v$ compares well with the experimental data and is well-reproduced by the ADP and MT potentials. The ADP potential slightly overestimates the DFT value while the MT potential underestimates it, but both predictions are in reasonable agreement with DFT considering the uncertainty of the calculations. In contrast, the EAM1 and EAM2 potentials significantly underestimate the migration energy. A more detailed comparison is presented in Fig.~\ref{fig:neb} showing the energy variation along the vacancy migration path. 

Self-interstitials in FCC metals typically form a dumbbell configuration with a $\langle 100 \rangle$, $\langle 110 \rangle$, or $\langle 111 \rangle$ orientation. In Table \ref{table:table1}, these dumbbells are denoted d100, d110, and d111, respectively. The table also includes localized interstitials that occupy an octahedral or tetrahedral site (${\rm Int}_{\rm oct}$ and ${\rm Int}_{\rm tet}$, respectively). In most  FCC metals, the $\langle 100 \rangle$ dumbbell orientation is the most energetically favorable interstitial configuration. However, our DFT calculations show that an octahedral interstitial has even lower energy. The difference is significant (0.35 eV), indicating that Pt is an exception to this rule. This observation aligns with recent DFT calculations \cite{Ma:2021aa} showing that octahedral interstitials are more stable than $\langle 100 \rangle$ dumbbells in Pt, Rh, and Th. For Pt, the difference was found to be 0.33 eV with a PBE functional and 0.36 with a PBEsol functional \cite{Ma:2021aa}. Both numbers agree well with our calculations. In contrast, all potentials tested here predict the  $\langle 100 \rangle$ dumbbell to be the ground state. This is unsurprising because the mentioned anomaly is caused by subtle electronic effects that are unlikely to be captured by classical potentials. We also note that the EAM1 and EAM2 potentials systematically underestimate the interstitial energies, while the ADP potential overestimates them. 

The unreconstructed surface energies predicted by the potentials follow the same ranking as in the DFT calculations, i.e., $\gamma_s(111) < \gamma_s(100) < \gamma_s(110)$, which is consistent with the experimental (semi-empirical) data \cite{Kim:2018vb}. Note that the EAM1 potential consistently over-predicts the DFT  surface energies. It is known that the unreconstructed (110) and (100) surfaces in pure Pt are unstable against reconstructions into lower-energy structures. In particular, the most energetic (110) surface reconstructs into several structures, the best-known of them being  the $(1\times2)$ and $(1\times3)$ missing-row reconstructions observed experimentally \cite{Sowa:1988aa,Robinson:1993aa} and verified by DFT calculations \cite{Chen:2011ab,Schultz:2021aa}. Our DFT calculations show that the energies of these two reconstructed surfaces are close to each other and significantly lower than the unreconstructed surface energy, in agreement with previous reports. In contrast, none of the four potentials correctly predicts the reconstructions. In all cases, the reconstructed surface energy is substantially higher than that of the unreconstructed surface. This is a common deficiency of the four potentials. In contrast, the previous EAM \cite{Foiles86,Adams89a} and MEAM \cite{Lee03a} potentials reproduced the negative sign of the energy change that accompanied the $(1\times2)$ reconstruction of the (110) surface, although the magnitude of the energy difference was significantly lower than predicted by the DFT calculations (Table \ref{table:table_classical_potential}).  The recently developed machine-learning potential in the GAP format GAP \cite{Kloppenburg:2023aa} also reproduces this reconstruction (Table \ref{table:table_ML_potential}).

\begin{figure}[ht]
\vspace{0.0cm}
\centering
\includegraphics[height=6.5cm]{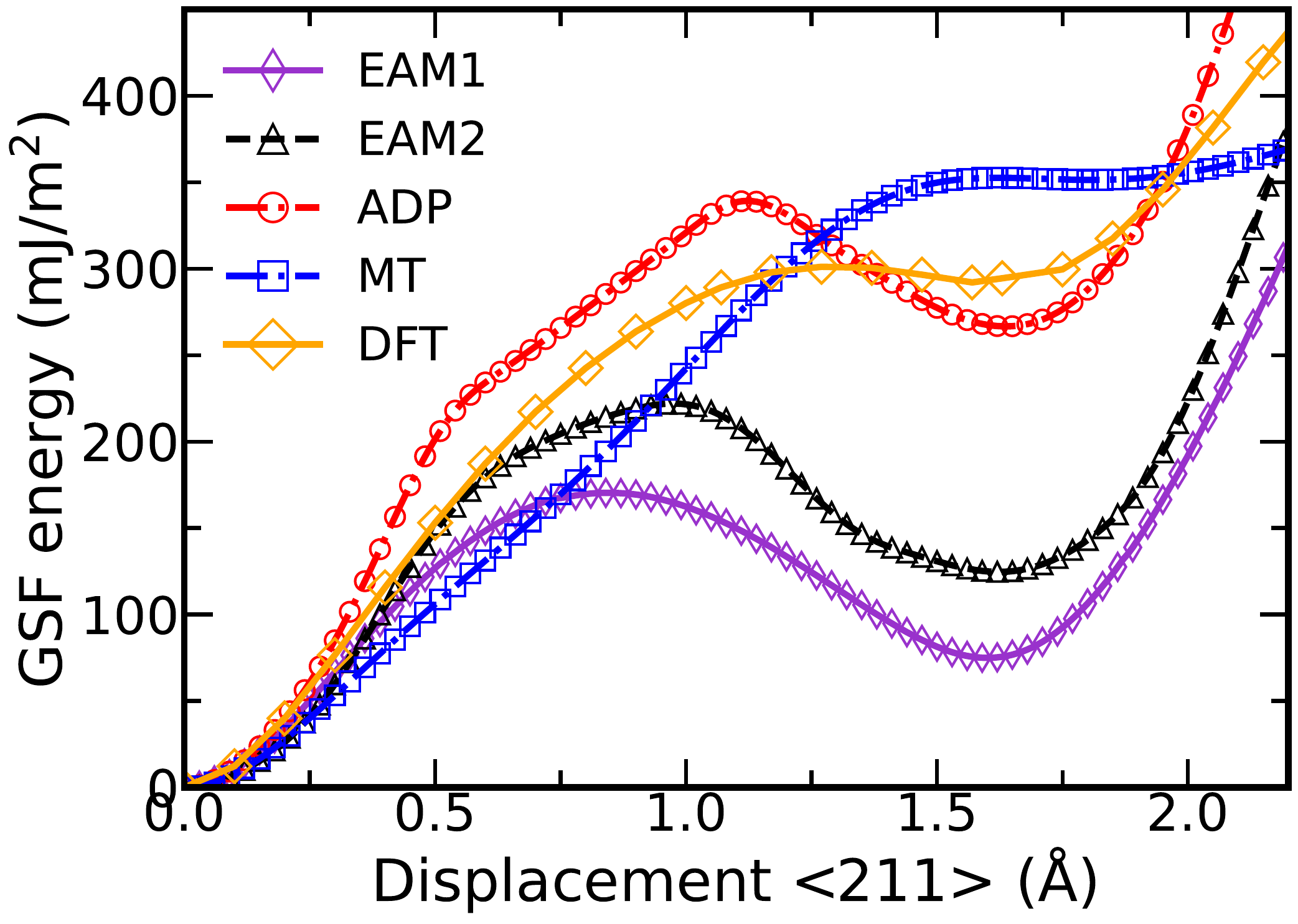}
\caption[]{Generalized stacking (GSF) fault energy as a function of displacement  in  Pt obtained by DFT calculations in comparison with interatomic potentials.}
\label{fig:SFE}
\end{figure}

The most significant differences among the four potentials are revealed when testing their ability to reproduce the twin boundary energy ($\gamma_{\rm twin}$), the intrinsic stacking fault energy (SFE) $\gamma_{\rm SF}$, and the unstable stacking fault energy (USFE) $\gamma_{\rm USF}$. The ADP potential reproduces $\gamma_{\rm twin}$ in excellent agreement with the DFT value (142 mJ\medspace{m$^{-2}$}), the MT potential overestimates it by 50 mJ\medspace{m$^{-2}$}, while the EAM1 and EAM2 potentials drastically underestimate the DFT calculation. For example, the EAM1 potential yields $\gamma_{\rm twin} = 46\ {\rm mJ}\medspace{{\rm m}^{-2}}$. The stacking fault energies were determined from the minima and maxima of the $\gamma$-surface; see Figure \ref{fig:SFE}. Note the significant difference between the SFE  predicted by our DFT calculation (287 mJ\medspace{m$^{-2}$}) and the experimental estimate (322 mJ\medspace{m$^{-2}$}) \cite{Murr,anderson2017theory}. Our DFT value is consistent with previous calculations; e.g.,  282 mJ\medspace{m$^{-2}$} \cite{Wu:2010ab}, 284 mJ\medspace{m$^{-2}$} \cite{Yao:2021aa}, and 286 mJ\medspace{m$^{-2}$} \cite{Jin:2011aa}. Considering the assumptions and approximations involved in the experimental SFE estimates, especially when the SFE is large, we suggest that the results obtained by DFT calculations are more reliable. 

As indicated in Table \ref{table:table1}, the SFE predicted by the ADP potential is in the closest agreement with the DFT calculation. The MT potential overshoots the DFT value by 23\%. The EAM1 and EAM2 potentials grossly underestimate the DFT calculation, especially the EAM1 potential, which predicts $\gamma_{\rm SF} = 74$ mJ\medspace{m$^{-2}$}. A similar trend is observed for the USFE. The ADP and MT potentials over-predict the DFT value of USFE while the EAM1 and EAM2 potentials significantly under-predict it. 

\subsection{Structural stability of Pt}

All four potentials  correctly predict that the FCC structure has a lower cohesive energy $E_0$ than all other structures tested in this work (Table \ref{table:table1}). The potentials also correctly place the HCP structure slightly above FCC, and the BCC structure slightly above HCP. The remaining crystal structures are  higher in energy and play a lesser role in atomistic simulations. However, it was interesting to compare the potential predictions with DFT calculations as a measure of their transferability to highly non-equilibrium environments. Since potentials and DFT calculations yield slightly different values of $E_0$, a comparison was made for the energy differences between the alternate structural energies and $E_0$. 

\begin{figure}[ht]
\vspace{0.0cm}
\centering
\includegraphics[height=7cm]{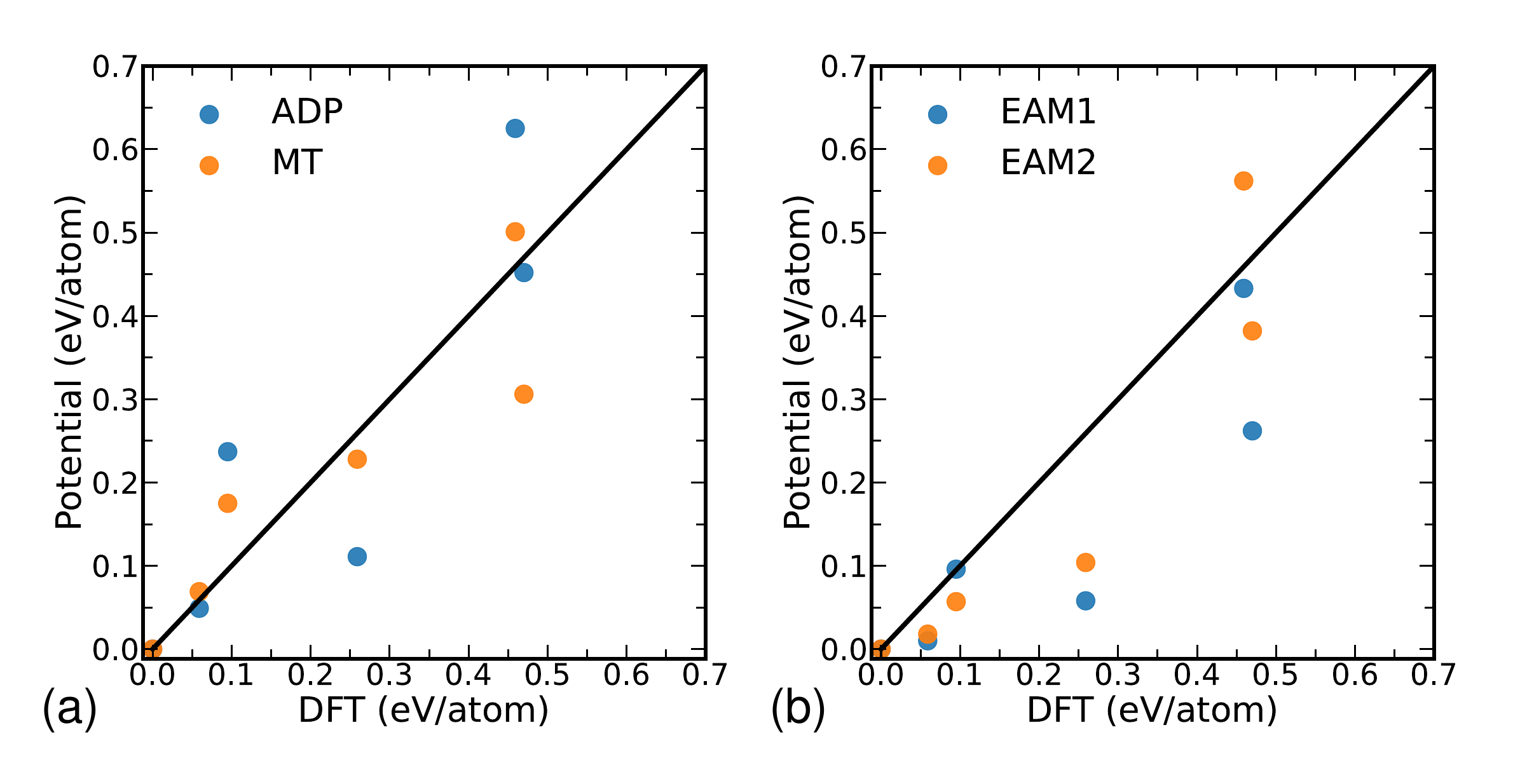}
\caption[]{Energies of crystal structures of Pt relative to the equilibrium FCC structure computed with interatomic potentials and obtained by DFT calculations. (a) ADP and MT potentials developed in this work. (b) EAM1 \cite{Zhou2004a} and EAM2 \cite{OBrien:2018aa} potentials. The straight line represents perfect correlation.}
\label{fig:structures}
\end{figure}

In Fig.~\ref{fig:structures}, we show  parity plots of the relative energies computed with the potentials against those obtained by DFT calculations. Each structural energy was equilibrated with respect to  volume and shape deformation. The plots consider only the FCC, BCC, A15, HEX, SC, and SH (simple hexagonal) structures. The DC and dimer structures are not included since their energies are too high and are hardly relevant to this comparison. All four potentials display a large scatter of the points while still capturing the general trend. However, note that for the ADP and MT potentials, the scatter is unbiased, whereas the EAM1 and EAM2 potentials show a clear trend towards underestimated DFT energies. 

\section{Discussion and conclusions\label{sec:conclusions}}
We have developed two new interatomic potentials for Pt by fitting the potential parameters to a database of DFT energies. No experimental data has been used during potential training. One of the potentials is in the well-established ADP format \citep{Mishin05a,Mishin.Ta,Hashibon08,Apostol2010,Apostol2011}, an extension of EAM to include non-central atomic interactions. The other potential is in the less conventional MT format. The latter is an extension of the covalent MT  format ~\citep{Kumagai:2007ly,Purja-Pun:2017aa} to metallic systems by significantly expanding the cutoff sphere of atomic interactions.  

The capabilities of the new potentials have been evaluated by comparing them with two of the recently published and probably most reliable EAM potentials   \cite{Zhou2004a,OBrien:2018aa}. Extensive tests have shown that the new ADP and MT potentials are more accurate than the EAM potentials. In particular, they cure some of the deficiencies of the EAM potentials, such as the underestimated stacking fault energies, the vacancy migration energy, and the phonon frequencies, as well as the grossly under-predicted melting temperature.

Interestingly and perhaps somewhat surprisingly, the MT  potential is at least as accurate as the ADP potential and even outperforms it for some properties (see Table \ref{table:table1}). The recently developed MT potential for Al has also been found to be at least as accurate as EAM potentials \cite{Stacking-faults_arxiv}. Remarkably, MT potentials have only 16 fitting parameters compared to 36 in the  ADP Pt potential. This makes the MT model more efficient than ADP. On the downside, MT potentials are computationally slow. For example, MD simulations with the MT potential are over a factor of $10^2$ slower than with the ADP potential. The cause of this slow performance is that the MT model includes an explicit treatment of the bond angles (section \ref{sec:MT}). Their calculation is performed in two nested loops over neighbors, whereas the EAM/ADP potentials use a single loop. The double-loop calculation  does not cause a significant slowdown for covalent materials, for which only the first (or sometimes also the second) neighbors are considered. However, when dozens of neighbors are included, the bond-angle calculation becomes the computational bottleneck.

\begin{figure}[ht]
\vspace{0.0cm}
\centering
\includegraphics[height=6.5cm]{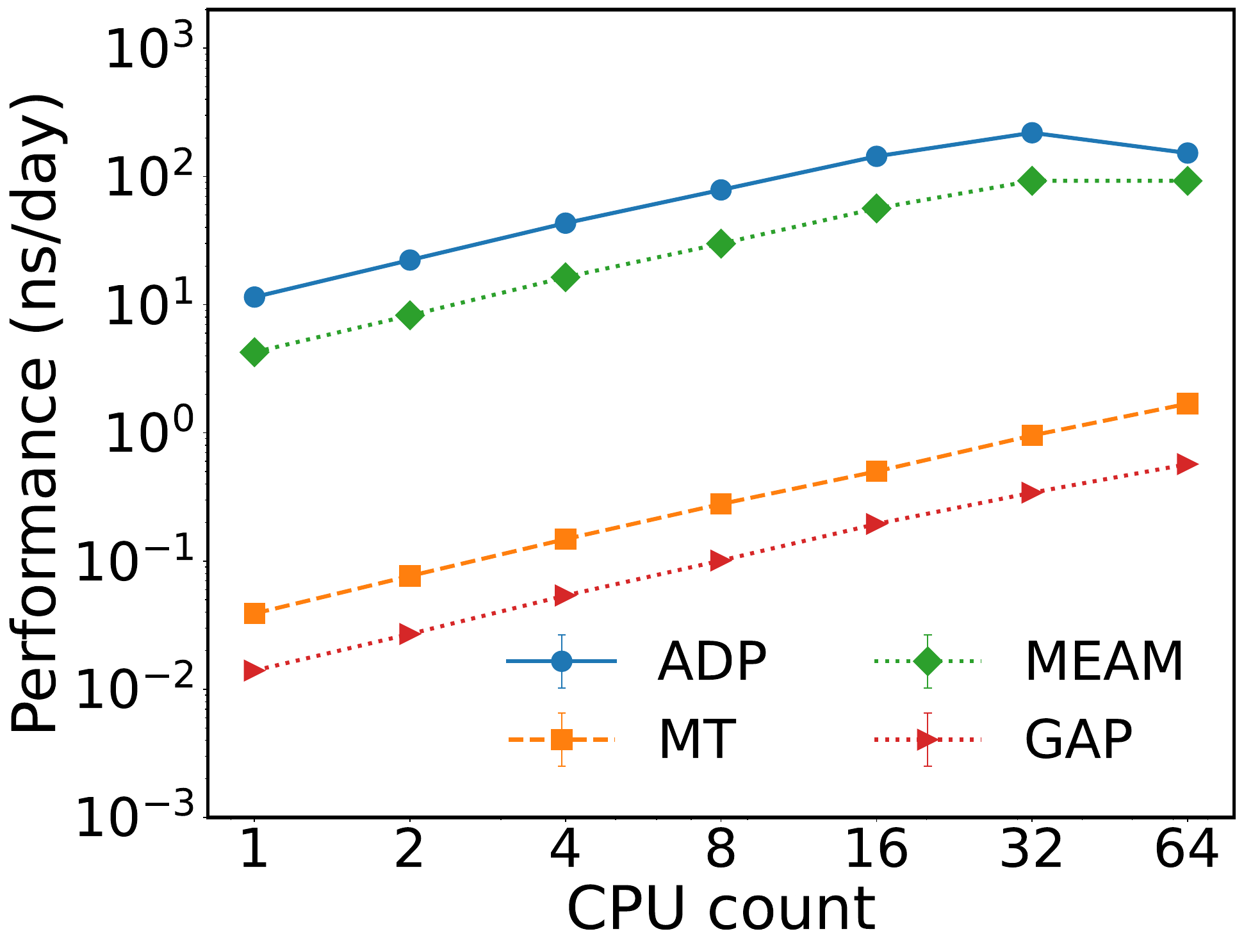}
\caption[]{Computational speed comparison of the interatomic potentials for Pt tested in this work. The ADP and MT potentials were developed in this work, while MEAM \cite{Lee03a} and GAP \cite{Kloppenburg:2023aa} potentials were reported in the literature.}
\label{fig:Speed}
\end{figure}

Figure \ref{fig:Speed} presents a more detailed comparison of the computational performance of Pt potentials, including the two new potentials developed in this work and the MEAM \cite{Lee03a} and GAP \cite{Kloppenburg:2023aa} potentials from the literature. The plot shows the number of nanoseconds per day during MD simulations of a solid-liquid system ($10^4$ atoms) using different numbers of CPUs ranging from 1 to 64. The tests are based on the implementation of these potentials in LAMMPS.
The ADP potential exhibits the highest computational efficiency. The computational speeds of the MEAM, MTP, and GAP potentials relative to ADP are approximately 1/3, 1/290, and 1/820, respectively. With an increasing number of CPUs, the speed gap between the ADP/MEAM and MT/GAP potentials tends to narrow due to the good scaling of the parallel implementation of the MT and GAP potentials in LAMMPS. 

Given the large computational overhead of the MT potentials, one may question their practical relevance. However, the MT model may possibly fill a critical gap between metallic potentials such as EAM, MEAM and ADP, and purely covalent potentials in the Tersoff \citep{Tersoff88,Tersoff:1988dn,Tersoff:1989wj} and Stillinger-Weber \cite{Stillinger85} formats. Many technological materials are dominated by hybrid interatomic bonding combining some features of the metallic and covalent interactions. Such materials include metallic carbides, borides, nitrides, and silicides, as well as heterophase systems containing metal/nonmetal interfaces. Developing classical physics-based potentials for such systems is a formidable challenge. Neither purely metallic nor short-range covalent potentials are capable of representing the mixed-bonding chemistry. These two classes of potentials have different and largely incompatible analytical forms, which makes the attempts to mix them together highly challenging \cite{Saidi_2014}. MT potentials might offer a solution due to their ability to represent both metals and covalent materials within the \emph{same} analytical model. It should also be mentioned that an MT potential for a binary system is a flexible, parameter-rich model containing up to $\approx 10^2$ free parameters. These advantages might outweigh the high computational cost of MT potentials.

Specifically for Pt, the MT Pt potential developed in this work could be crossed with short-range covalent MT potentials for elements such as C, Se, Si or N. For example, platinum silicides, such as PtSi and Pt$_2$Si, play an important role in microelectronic device technology \cite{Chen:2005aa,Taylor:2021aa}.
Pt-N coordination compounds are commonly used in platinum-based anticancer drugs \cite{Johnstone:2014aa}. Reliable interatomic potentials for these and similar systems would enable large-scale atomic-level computer simulations of their mechanical and physical behaviors and chemical interactions across interfaces. 

\bigskip{}

\noindent\textbf{Acknowledgments}

\noindent This research was supported by the National Institute of Standards and Technology, Material Measurement Laboratory, the Materials Science and Engineering Division.


\newpage{}

\begin{table}[!h]
\centering{}{\small\caption[]{DFT database  used for the development of the Pt potentials. $N_{A}$ is the number of atoms per supercell and $N_{tv}$ is the number of supercells in the subset.} 
}\bigskip{}
{\small{} }{\small{}%
\begin{tabular}{|c|c|c|c|c|}
\hline 
{\small Subset} & {\small Structure} & {\small Simulation type} & {\small $N_{A}$} & {\small $N_{tv}$}\tabularnewline
\hline 
{\small Crystal EOS} & {\small FCC} & {\small Isotropic strain } & {\small 4} & {\small 333}\tabularnewline
\hline 
 & {\small SC} & {\small Isotropic strain } & {\small 1} & {\small 281}\tabularnewline
\hline 
 & {\small DC} & {\small Isotropic strain } & {\small 8} & {\small 574}\tabularnewline
\hline 
 & {\small BCC} & {\small Isotropic strain} & {\small 2} & {\small 366}\tabularnewline
\hline 
 & {\small A15} & {\small Isotropic strain} & {\small 8} & {\small 483}\tabularnewline
\hline 
 & {\small HCP} & {\small Isotropic strain} & {\small 4} & {\small 99}\tabularnewline
\hline 
 & {\small HCP} & {\small Perturbation on isotropic strain } & {\small 4} & {\small 101}\tabularnewline
\hline 
 &  &  &  & \tabularnewline
\hline 
{\small Deformation} &  & {\small Uniaxial strain along $\langle$100$\rangle$ } & {\small 8} & {\small 51}\tabularnewline
\hline 
 &  & {\small Uniaxial strain along $\langle$100$\rangle$ (different size)} & {\small 4} & {\small 33}\tabularnewline
\hline 
 &  & {\small Uniaxial strain along $\langle$110$\rangle$} & {\small 4} & {\small 51}\tabularnewline
\hline 
 &  & {\small Uniaxial strain along $\langle$111$\rangle$ } & {\small 24} & {\small 51}\tabularnewline
\hline 
 &  & {\small Shear preserving volume ($C_{11}-C_{12}$)} & {\small 4} & {\small 21}\tabularnewline
\hline 
 &  & {\small Shear non-preserving volume ($C_{44}$)} & {\small 4} & {\small 101}\tabularnewline
\hline 
 &  &  &  & \tabularnewline
\hline 
{\small Compression} &  & {\small Strong isotropic compression } & {\small 32} & {\small 216}\tabularnewline
\hline 
 &  & {\small Strong uniaxial compression along $\langle$100$\rangle$ } & {\small 32} & {\small 151}\tabularnewline
\hline 
 &  & {\small Strong uniaxial compression along $\langle$110$\rangle$ } & {\small 48} & {\small 99}\tabularnewline
\hline 
 &  & {\small Strong uniaxial compression along $\langle$111$\rangle$ } & {\small 36} & {\small 50}\tabularnewline
\hline 
 &  &  &  & \tabularnewline
\hline 
{\small Surface} & {\small (100)} & {\small (100) surface at high temperature } & {\small 8} & {\small 30}\tabularnewline
\hline 
 & {\small (110)} & {\small (110) surface at high temperature } & {\small 16} & {\small 21}\tabularnewline
\hline 
 & {\small (111)} & {\small (111) surface at high temperature } & {\small 9} & {\small 33}\tabularnewline
\hline 
 & {\small (210)} & {\small (111) surface at high temperature } & {\small 20} & {\small 27}\tabularnewline
\hline 
 & {\small (221)} & {\small (111) surface at high temperature } & {\small 18} & {\small 10}\tabularnewline
\hline 
 &  &  &  & \tabularnewline
\hline 
{\small Migration barrier} &  & {\small Minimum energy path (DFT NEB)} & {\small 31} & {\small 9}\tabularnewline
\hline 
 &  & {\small Maximum energy path } & {\small 31} & {\small 45}\tabularnewline
\hline 
 &  &  &  & \tabularnewline
\hline 
{\small Liquid} &  & {\small DFT evaluated liquid snapshots} & {\small 32} & {\small 101}\tabularnewline
\hline 
 &  &  &  & \tabularnewline
\hline 
{\small Dimer} &  & {\small Isotropic strain } & {\small 2} & {\small 101}\tabularnewline
\hline 
\end{tabular}}{\small\label{table:DFT-database}}
\end{table}

\begin{table}
\caption{Optimized parameters of the ADP potential for Pt developed in this
work.\label{tab:Optimized-parameters-ADP}}

\bigskip{}

\centering{}%
\begin{tabular}{|c|c|c|c|c|}
\hline 
Parameter & Value &  & Parameter & Value\tabularnewline
\hline 
\hline 
$r_{c}$~($\textrm{\AA}$) & $0.5497453\times10^{1}$ &  & $r_{05}$~($\textrm{\AA}$) & $0.7166108\times10^{0}$\tabularnewline
\hline 
$h$~($\textrm{\AA}$) & $0.3722373\times10^{1}$ &  & $b_{5}$~($\textrm{\AA}$$^{-1}$) & $-0.1760697\times10^{-1}$\tabularnewline
\hline 
$E_{1}$ (eV) & $0.7215604\times10^{2}$ &  & $E_{6}$ (eV) & $0.1679562\times10^{-1}$\tabularnewline
\hline 
$r_{01}$~($\textrm{\AA}$) & $0.2317099\times10^{1}$ &  & $r_{06}$~($\textrm{\AA}$) & $0.2566572\times10^{0}$\tabularnewline
\hline 
$b_{1}$~($\textrm{\AA}$$^{-1}$) & $0.5400912\times10^{0}$ &  & $b_{6}$~($\textrm{\AA}$$^{-1}$) & $-0.4611701\times10^{-1}$\tabularnewline
\hline 
$E_{2}$ (eV) & $-0.4558172\times10^{-2}$ &  & $s_{3}$ (eV) & $0.2979066\times10^{-1}$\tabularnewline
\hline 
$r_{02}$~($\textrm{\AA}$) & $0.3334695\times10^{1}$ &  & $E_{7}$ (eV) & $-0.3690955\times10^{-1}$\tabularnewline
\hline 
$b_{2}$~($\textrm{\AA}$$^{-1}$) & $-0.3401386\times10^{1}$ &  & $r_{07}$~($\textrm{\AA}$) & $0.1240603\times10^{0}$\tabularnewline
\hline 
$s_{1}$ (eV) & $0.2283578\times10^{2}$ &  & $b_{7}$~($\textrm{\AA}$$^{-1}$) & $-0.1288544\times10^{-1}$\tabularnewline
\hline 
$m$ (eV) & $0.2811670\times10^{2}$ &  & $E_{8}$ (eV) & $0.1666801\times10^{-2}$\tabularnewline
\hline 
$E_{3}$ (eV) & $0.7372560\times10^{3}$ &  & $r_{08}$~($\textrm{\AA}$) & $-0.8520997\times10^{-1}$\tabularnewline
\hline 
$r_{03}$~($\textrm{\AA}$) & $0.6283386\times10^{1}$ &  & $b_{8}$~($\textrm{\AA}$$^{-1}$) & $-0.1547745\times10^{-1}$\tabularnewline
\hline 
$b_{3}$~($\textrm{\AA}$$^{-1}$) & $-0.1671072\times10^{1}$ &  & $s_{4}$ (eV) & $-0.1314255\times10^{0}$\tabularnewline
\hline 
$E_{4}$ (eV) & $-0.1182929\times10^{0}$ &  & $\rho_{0}$ & $-0.2168302\times10^{-1}$\tabularnewline
\hline 
$r_{04}$~($\textrm{\AA}$) & $0.1258595\times10^{2}$ &  & $a$ & $-0.6253520\times10^{2}$\tabularnewline
\hline 
$b_{4}$~($\textrm{\AA}$$^{-1}$) & $-0.7552921\times10^{-1}$ &  & $b$ & $0.1886500\times10^{2}$\tabularnewline
\hline 
$s_{2}$ (eV) & $0.1806989\times10^{1}$ &  & $c$ & $-0.5358983\times10^{1}$\tabularnewline
\hline 
$E_{5}$ (eV) & $-0.4166797\times10^{0}$ &  & $d$ & $0.2068855\times10^{-1}$\tabularnewline
\hline 
\end{tabular}
\end{table}

\begin{table}
\centering{}\caption{Optimized parameters of the MT potential for Pt developed in this
work.\label{tab:Optimized-parameters-MT}}
\bigskip{}
\begin{tabular}{|l|c|c|c|c|}
\hline 
Parameter & Value &  &  & \tabularnewline
\hline 
$\ln\left[A(\mathrm{eV})\right]$ & $7.997930\times10^{0}$ &  & $c_{1}$ & $9.223200\times10^{-3}$\tabularnewline
\hline 
$\ln\left[B(\mathrm{eV})\right]$ & $4.812509\times10^{0}$ &  & $c_{2}$ & $2.354200\times10^{-2}$\tabularnewline
\hline 
$\lambda_{1}$ (\AA$^{-1}$) & $2.921020\times10^{0}$ &  & $c_{3}$ & $2.627790\times10^{-2}$\tabularnewline
\hline 
$\lambda_{2}$ (\AA$^{-1}$) & $1.527300\times10^{0}$ &  & $c_{4}$ & $2.414760\times10^{0}$\tabularnewline
\hline 
$\eta$ & $6.525730\times10^{0}$ &  & $c_{5}$ & $7.079140\times10^{0}$\tabularnewline
\hline 
$\eta\times\delta$ & $5.311480\times10^{0}$ &  & $h$ & $6.63771\times10^{-1}$\tabularnewline
\hline 
$\alpha$ & $2.961940\times10^{-1}$ &  & $R_{1}$ (\AA) & $5.587870\times10^{0}$\tabularnewline
\hline 
$\beta$ & $1$ &  & $R_{2}$ (\AA) & $6.566095\times10^{0}$\tabularnewline
\hline 
$c_{0}$ (eV) & $-1.748910\times10^{-2}$ &  &  & \tabularnewline
\hline 
\end{tabular}
\end{table}

\begin {table} [h]
  \center
  \caption {Pt properties predicted by interatomic potentials in comparison with DFT calculations performed in this work and experimental data. The potentials EAM1 \cite{Zhou2004a} and EAM2 \cite{OBrien:2018aa} were taken from the literature while the ADP and MT potentials were developed in this work. $^{*}$ Recomputed value differs from the one reported in  \cite{OBrien:2018aa}.  The properties we consider especially inaccurate are typest in bold.}   
  \begin{tabular}{l cl cl cl cl cl  cl c|}
   \hline	
Properties     		 		 &      EAM1        	   	&    EAM2    	&	ADP			&            MT                       & DFT  &  Experiment      \\       \hline                                                  
$a_0$ ({\AA})	    			 &       3.920	  	 	&    3.976        		&	4.010	&	      3.971       		&   3.967        &       3.92$^{b,i}$     \\                                                                   
$E_{0}$ (eV/atom)		&       \(-5.770\)      		&    \(-5.506\)      	 	&	\(-5.841\)	&           \(-5.845\)   		&    \(-5.931\)              &  \(-5.84\)$^i$   \\                                                                             
$B$   (GPa)   	       			 &       283.0	   	 	&    236.1         		&	283.8	&	      249.6         		&   253.2         &     288$^{h}$,  289$^{f}$       \\                                                         
$C_{11}$ (GPa)      	        		 	&       347.2	     		&    264.3        		&	356.9 	&          295.3       &   323.1           &       348.7$^{g}$    \\                                                                   
$C_{12}$ (GPa)      	        		 	&       250.9	     		&    221.9       		&	247.3	 &          226.8       &   218.2     &      245.8$^{g}$       \\                                                                  
$C_{44}$ (GPa)  	        		 	&       76.5      		         &    68.8      		&	80.4	 	&          74.4       &   67.8   &     73.4$^{g}$        \\                                                                  
$E^{v}_{f}$  (eV)    		 	&       1.53	     		         &    1.89     		&	1.48	&	         1.34      &   0.63            &   1.24-1.45$^{a}$ \\                                                   
$E_{m}^{v}$  (eV)      	 	&	 \textbf{0.73}	     		&    \textbf{ 0.84}           &	1.36	&	    		1.10           &      1.30                         &    1.38-1.45$^{j}$                      \\                                                  
d100 (eV)    	       	         	&       3.91    		 &    3.89      &	6.01	&	         4.47       &        5.32        &                                     \\                                                          
d110 (eV)    	        			&       4.23     		&    4.15       &	6.67	&	        5.19       &       6.01         &                                     \\                                                          
d111 (eV)    	        			 &       4.38	    	 &    4.32      &	7.11	&	        5.69       &      6.00          &                                     \\                                                          
Int$_{\rm oct}$(eV) 	    			 &       4.12       		 &    4.12     &	6.19	&	          4.63       &    5.01            &                                     \\                                                          
Int$_{\rm tet}$ (eV) 	    			 &       4.45	     		&    4.35     &	7.23	&	          5.82       &      6.29          &                                    \\                                                          
$\gamma_{s}$(100)  (J m$^{-2}$)       	 	&       2.177	   		 &    1.740          &		1.994	          &          1.662                 &  1.824       &   2.036$^{c}$        \\                                                      
$\gamma_{s}$(110)   (J m$^{-2}$)         	 	&       2.396			&    1.828            &		2.126	        &          1.890                  &  1.862           &    2.106$^{c}$         \\                                                      
$\gamma_{s}$(111)  (J m$^{-2}$)       	 	&       2.082			 &    1.618          &		1.921	          &          1.434                   &  1.483        &    1.502$^{c}$        \\                                                      
$\gamma_{s}$(110)$_{\rm R}$ ($1\times2$)  (J m$^{-2}$)  	&       2.473			&    1.893            &	2.228	&	         1.991                 &     1.682             &                \\                                                   
$\gamma_{s}$(110)$_{\rm R}$ ($1\times3$)	 (J m$^{-2}$)	&	  2.480			 &    1.917            &	2.259	      		&   1.903		                                  &       1.660             &                 &                \\                                       
$\gamma_{\rm twin}$     (mJ m$^{-2}$)			&      \textbf{46.09}				 &    \textbf{66.9}$^{*}$                      &     140.9 &     	\textbf{190.9}                 &  142.1                                 &                \\                                                  
$\gamma_{\rm SF}$       (mJ m$^{-2}$)		& \textbf{74.3}	  &    \textbf{124.6}$^{*}$      &	266.6	&	352.6 &                             287.3              &      322$^{e}$          \\                                                               
$\gamma_{\rm USF}$       (mJ m$^{-2}$)		&	\textbf{170.6}		   &    222.0   			&         339.2                    &  352.6     &  301.1 &             &                \\                                                               
HCP    (eV/atom)  		&	-5.760       &    -5.488     & -5.792	&	     	-5.776    &   -5.872    &                                     \\                                                         
BCC  (eV/atom)    		&	-5.674       &    -5.449     &	-5.604	&	          -5.670   &   -5.836    &                                       \\                                                         
SC     (eV/atom)  		&	-5.337       &    -4.944     &	-5.216	&	          -5.344   &    -5.472    &                                        \\                                                         
DC (eV/atom)  		&	-4.423       &    -4.177     &	-4.337	&	          -3.280   &   -4.855      &                                        \\                                                        
SH      (eV/atom) 		&	-5.508       &    -5.124     &	-5.389	&	          -5.539   &    -5.461        &                                        \\                                                         
A15    (eV/atom)  		&	-5.712       &    -5.402     &	-5.730	&	          -5.617   &    -5.672   &                                        \\                                                         
Dimer  (eV/atom)  		&	-0.199       &    -2.679     &	-1.117	&	         -0.448    &     -2.065        &                                        \\       
$T_{m}$ (K)				&	\textbf{1474}  &     \textbf{1784}   &	 2041		&        	2195                               &                 &                \(2045\)$^{i}$        \\                                                        
$\nu	$$_{\rm max}$ (THz) 		&	\textbf{4.88}	  	  &     \textbf{4.37}           &	5.20 	&	        	5.35      &        5.56         &      5.85$^{d}$                                  \\                                                        
\hline	
\end{tabular}   

$^{a}$Ref.~\cite{Mattsson:2002},  $^{b}$Ref.~\cite{Miller:2009}, $^{c}$Ref.~\cite{Kim:2018vb}, $^{d}$Ref.~\cite{dutton1972crystal}, $^{e}$Ref.~\cite{Murr}, $^{f}$Ref.~\cite{Nazarov:2012aa}, $^{g}$Ref.~\cite{Kamada:2019wc}, $^{h}$Ref.~\cite{macfarlane1965}
$^{i}$Ref.~\cite{Kittel}, $^{j}$Ref.~\cite{Balluffi78}

 \label{table:table1}
\end{table}

\newpage \clearpage

\newpage \clearpage 

\newpage \clearpage 

\section{Appendix}

\begin{table}[h]
  \center
  \caption{Pt properties computed with interatomic potentials in comparison with DFT calculations and experimental data. The potentials labeled as MEAM \cite{Lee03a}, EMT (effective medium theory) \cite{Jacobsen1996}, EAM3 \cite{Adams89a}, and EAM4 \cite{Foiles86} were downloaded from the NIST Interatomic Potential Repository at (https://www.ctcms.nist.gov/potentials/). The properties we consider especially inaccurate are typest in bold. \label{tab:other_EAM}}
    \bigskip{}
  \begin{tabular}{l c c c c c c}
   \hline
Properties & MEAM & EMT & EAM3 & EAM4 & DFT & Experiment \\ \hline
$a_0$ ({\AA}) & 3.917 & 3.922 & 3.920 & 3.920 & 3.967 & 3.92$^{b,i}$ \\
$E_{0}$ (eV/atom) & \(-5.770\) & \(-5.850\) & \(-5.770\) & \(-5.770\) & \(-5.931\) & \(-5.84\)$^i$ \\
$B$ (GPa) & 288.4 & 278.7 & 283.1 & 283.1 & 253.2 & 288$^{h}$, 289$^{f}$ \\
$C_{11}$ (GPa) & 358.1 & 318.6 & 304.1 & 303.1 & 323.1 & 348.7$^{g}$ \\
$C_{12}$ (GPa) & 253.5 & 258.7 & 272.6 & 273.1 & 218.2 & 245.8$^{g}$ \\
$C_{44}$ (GPa) & 77.5 & 79.4 & 71.9 & 68.3 & 67.8 & 73.4$^{g}$ \\
$E^{v}_{f}$ (eV) & 1.50 & 1.02 & 1.77 & 1.69 & 0.63 & 1.24--1.45$^{a}$ \\
$E_{m}^{v}$ (eV) & 1.43 & 0.98 & \textbf{0.86} & \textbf{0.82} & 1.30 & 1.38--1.45$^{j}$ \\
d100 (eV) & 6.75 & 4.95 & \textbf{3.43} & \textbf{3.29} & 5.36 & \\
d110 (eV) & 7.75 & 5.23 & \textbf{3.55} & \textbf{3.39} & 6.01 & \\
d111 (eV) & \bf{4.38} & 5.69 & \textbf{3.58} & \textbf{3.42} & 6.00 & \\
Int$_{\rm oct}$ (eV) & 7.51 & 5.03 & \textbf{3.64} & \textbf{3.48} & 5.01 & \\
Int$_{\rm tet}$ (eV) & 9.07 & 5.78 & \textbf{3.61} & \textbf{3.46} & 6.29 & \\
$\gamma_{s}$(100) (J m$^{-2}$) & 2.272 & \textbf{0.876} & 1.710 & 1.651 & 1.824 & 2.036$^{c}$ \\
$\gamma_{s}$(110) (J m$^{-2}$) & 2.187 & \textbf{0.931} & 1.823 & 1.754 & 1.862 & 2.106$^{c}$ \\
$\gamma_{s}$(111) (J m$^{-2}$) & 1.708 & \textbf{0.749} & 1.498 & 1.442 & 1.483 & 1.502$^{c}$ \\
$\gamma_{s}$(110)$_{\rm R}$ ($1\times2$) (J m$^{-2}$) & 2.047 & \textbf{0.905} & 1.792 & 1.721 & 1.682 & \\
$\gamma_{s}$(110)$_{\rm R}$ ($1\times3$) (J m$^{-2}$) & 1.993 & \textbf{0.904} & 1.794 & 1.723 & 1.660 & \\
$\gamma_{\rm twin}$ (mJ m$^{-2}$) & \textbf{55.3} & \textbf{2.5} & \textbf{7.6} & \textbf{7.4} & 142.1 & \\
$\gamma_{\rm SF}$ (mJ m$^{-2}$) & \textbf{110.3} & \textbf{5.0} & \textbf{15.1} & \textbf{14.6} & 287.3 & 322$^{e}$ \\
$\gamma_{\rm USF}$ (mJ m$^{-2}$) & \textbf{566.0} & \textbf{157.5} & \textbf{143.1} & \textbf{136.0} & 301.1 & \\
HCP (eV/atom) & -5.748 & -5.849 & -5.767 & -5.767 & -5.872 & \\
BCC (eV/atom) & -5.494 & -5.795 & -5.749 & -5.750 & -5.836 & \\
SC (eV/atom) & -5.005 & -5.386 & -5.209 & -5.237 & -5.472 & \\
DC (eV/atom) & -4.063 & -4.971 & -4.370 & -4.438 & -4.855 & \\
SH (eV/atom) & -5.460 & -5.571 & -5.435 & -5.452 & -5.461 & \\
A15 (eV/atom) & -5.591 & -5.714 & -5.701 & -5.704 & -5.672 & \\
Dimer (eV/atom) & -2.389 & -4.152 & -3.866 & -3.952 & -2.065 & \\
$T_{m}$ (K) & \textbf{2368} & \textbf{1355} & \textbf{1622} & \textbf{1528} & & \(2045\)$^{i}$ \\
$\nu_{\rm max}$ (THz) & 5.68 & \textbf{4.46} & \textbf{4.38} & \textbf{4.26} & 5.56 & 5.85$^{d}$ \\
\hline
\end{tabular}

$^{a}$Ref.~\cite{Mattsson:2002}, $^{b}$Ref.~\cite{Miller:2009}, $^{c}$Ref.~\cite{Kim:2018vb}, $^{d}$Ref.~\cite{dutton1972crystal}, $^{e}$Ref.~\cite{Murr}, $^{f}$Ref.~\cite{Nazarov:2012aa}, $^{g}$Ref.~\cite{Kamada:2019wc}, $^{h}$Ref.~\cite{macfarlane1965}, $^{i}$Ref.~\cite{Kittel}, $^{j}$Ref.~\cite{Balluffi78}

\label{table:table_classical_potential}
\end{table}

\begin{table}[h]
  \center
  \caption{Pt properties computed with the GAP potential \cite{Kloppenburg:2023aa} in comparison with DFT calculations and experimental data. The properties we consider especially inaccurate are typest in bold.\label{tab:gap}}
  \begin{tabular}{l c c c}
   \hline
Properties & GAP & DFT & Experiment \\ \hline
$a_0$ ({\AA}) & 3.968 & 3.967 & 3.92$^{b,i}$ \\
$E_{0}$ (eV/atom) & \(-6.098\) & \(-5.931\) & \(-5.84\)$^i$ \\
$B$ (GPa) & 246.7 & 253.2 & 288$^{h}$, 289$^{f}$ \\
$C_{11}$ (GPa) & 289.7 & 323.1 & 348.7$^{g}$ \\
$C_{12}$ (GPa) & 225.2 & 218.2 & 245.8$^{g}$ \\
$C_{44}$ (GPa) & 58.5 & 67.8 & 73.4$^{g}$ \\
$E^{v}_{f}$ (eV) & 0.83 & 0.63 & 1.24--1.45$^{a}$ \\
$E_{m}^{v}$ (eV) & 1.11 & 1.30 & 1.38--1.45$^{j}$ \\
d100 (eV) & 5.00 & 5.36 & \\
d110 (eV) & 5.51 & 6.01 & \\
d111 (eV) & 6.03 & 6.00 & \\
Int$_{\rm oct}$ (eV) & 5.43 & 5.01 & \\
Int$_{\rm tet}$ (eV) & 6.14 & 6.29 & \\
$\gamma_{s}$(100) (J m$^{-2}$) & 1.814 & 1.824 & 2.036$^{c}$ \\
$\gamma_{s}$(110) (J m$^{-2}$) & 1.840 & 1.862 & 2.106$^{c}$ \\
$\gamma_{s}$(111) (J m$^{-2}$) & 1.466 & 1.483 & 1.502$^{c}$ \\
$\gamma_{s}$(110)$_{\rm R}$ ($1\times2$) (J m$^{-2}$) & 1.755 & 1.682 & \\
$\gamma_{s}$(110)$_{\rm R}$ ($1\times3$) (J m$^{-2}$) & 1.708 & 1.660 & \\
$\gamma_{\rm twin}$ (mJ m$^{-2}$) & 97.6 & 142.1 & \\
$\gamma_{\rm SF}$ (mJ m$^{-2}$) & 238.9 & 287.3 & 322$^{e}$ \\
$\gamma_{\rm USF}$ (mJ m$^{-2}$) & 284.3 & 301.1 & \\
HCP (eV/atom) & -6.042 & -5.872 & \\
BCC (eV/atom) & -5.995 & -5.836 & \\
SC (eV/atom) & -5.676 & -5.472 & \\
DC (eV/atom) & -4.632 & -4.855 & \\
SH (eV/atom) & -5.821 & -5.461 & \\
A15 (eV/atom) & -5.860 & -5.672 & \\
Dimer (eV/atom) & -2.320 & -2.065 & \\
$T_{m}$ (K) & \textbf{1542} & & \(2045\)$^{i}$ \\
$\nu_{\rm max}$ (THz) & 5.50 & 5.56 & 5.85$^{d}$ \\
\hline
\end{tabular}

$^{a}$Ref.~\cite{Mattsson:2002}, $^{b}$Ref.~\cite{Miller:2009}, $^{c}$Ref.~\cite{Kim:2018vb}, $^{d}$Ref.~\cite{dutton1972crystal}, $^{e}$Ref.~\cite{Murr}, $^{f}$Ref.~\cite{Nazarov:2012aa}, $^{g}$Ref.~\cite{Kamada:2019wc}, $^{h}$Ref.~\cite{macfarlane1965}, $^{i}$Ref.~\cite{Kittel}, $^{j}$Ref.~\cite{Balluffi78}

\label{table:table_ML_potential}
\end{table}

\end{document}